# Al$_{5+\alpha}$Si$_{5+\delta}$N$_{12}$, a new Nitride compound

R. Dagher[1,3], L. Lymperakis[2], V. Delaye[3], L. largeau[4], A. Michon[1], J. Brault[1] and P. Vennéguès[1]

[1] Université Côte d'Azur, CRHEA-CNRS, rue B. Grégory, F-06560 Valbonne, France

[2] Max-Planck-Institut für Eisenforschung GmbH, Düsseldorf, Germany

[3] Université Grenoble Alpes, CEA, LETI, MINATEC Campus, F-38054 Grenoble, France

[4] C2N-CNRS, Route de Nozay, F-91460 Marcoussis, France

ABSTRACT

We report on the synthesis of new nitride-based compound by using annealing of AlN heteroepitaxial layers under a Si-atmosphere at temperatures between 1350°C and 1550°C. The structure and stoichiometry of this compound are investigated by high-resolution scanning transmission electron microscopy (HRSTEM), energy dispersive X-Ray (EDX) spectroscopy, and density functional theory (DFT) calculations. The identified structure is a derivative of the parent wurtzite AlN crystal where anion sublattice is fully occupied by N atoms and the cation sublattice is the stacking of 2 different planes along <0001>. The first one exhibits a ×3 periodicity along <10-10> with 1/3 of the sites being vacant. The rest of the sites in the cation sublattice are occupied by equal number of Si and Al atoms. Assuming a semiconducting alloy, which is expected to have a wide band gap, a range of stoichiometries is proposed, Al$_{5+\alpha}$Si$_{5+\delta}$N$_{12}$ with α being between 0 and 1/3 and δ between 0 and 1/4.

Due to their unique properties, wide-gap group III-Nitrides (III-N) materials are nowadays the materials of choice for optoelectronic applications especially for light emitting diodes which revolutionized domestic lightning in the last years. They also attracted more recently a lot of attention for high power and high frequency electronic applications. Silicon plays a major role for the properties of III-N. First, Si is the most efficient n-type dopant for III-N and is therefore extensively used for III-N devices. Moreover, the introduction of Si atoms may drastically modify the growth of III-N heteroepitaxial layers. In the case of GaN, the so called "SiN treatment" allows changing the growth mode from 2D to 3D[1,2,3,4]. The gradual formation of GaN pyramids leads to the bending of pre-existing dislocations and eventually to their mutual interaction and annihilation[5]. The "SiN treatment" is thus one of the most efficient methods for reducing the threading dislocation density in GaN heteroepitaxial layers, which is mandatory for improving the physical properties of the material up to device quality. The "SiN treatment", implemented during the metalorganic vapor phase epitaxial (MOVPE) growth of GaN, consists in exposing the growing surface for a few minutes to a silane/ammonia flux at temperatures above 1000°C. It was supposed that this process results in the formation of a $SiN_x$ micro-mask[1]. In fact, a detailed high-resolution transmission electron microscopy (TEM) study revealed the formation of a crystalline monolayer-thick $GaSiN_3$ compound epitaxied on GaN[6]. The atomic structure of this $GaSiN_3$ layer may be described in reference to the parent GaN wurtzite structure: the N-sublattice is preserved whereas the three cation positions in the proposed unit cell are occupied by one Si atom, one Ga atom and one vacancy, respectively. Thus, the exposure of the GaN surface to a silane/ammonia flux do not lead to the growth of an additional layer but to the replacement of surface Ga atoms by either Si atoms or vacancies. Such a $SiGaN_3$ monolayer buried by one monolayer of GaN acts as a dielectric mask and prohibits further GaN growth. Instead, the nucleation of 3D islands occurs in area non-covered by $GaSiN_3$.

The purpose of this work is to answer to several questions which arise from these results. The first one concerns the possibility of obtaining a SiGaN crystalline layer thicker than one monolayer, while previous results have shown that a long exposure of GaN surface to a silane/ammonia flux leads to the growth of an amorphous SiN layer[7]. On the other hand, it was reported that the in-situ passivation of AlGaN/GaN high electron mobility transistors leads to the growth of a crystalline $Si_3N_4$ layer[8,9]. The "SiN treatment" has already also been successfully implemented on AlN surfaces also resulting in a further 3D growth of GaN[10]. The second question therefore concerns the origin of this 3D growth on AlN and the possible existence of a SiAlN crystalline layer.

To tackle these questions our approach is to study the effect of the exposure of AlN surfaces to a silane flux at temperatures above 1300°C. In fact, it has been shown that at these temperatures atoms mobility is sufficiently high to promote crystalline quality improvement of AlN layers[11,12]. Such mobility may enhance the Al/Si atomic exchanges.

(0001)-oriented AlN heteroepitaxial films grown on sapphire either by molecular beam epitaxy (MBE)[13] or MOVPE[14] are annealed in a hot wall chemical vapor deposition (CVD) reactor in a silicon environment. As will be shown later, the following results do not depend on the growth technique of the AlN template layers. Two different processes are used, process 1 and process 2. Before process 1, the clean CVD reactor is exposed to a silane flux for 5 minutes at 1350°C and 200 mbars in a $H_2$ atmosphere. Then, process 1 consists in a simple annealing under $N_2$ flux at 1550°C and 800 mbars for few minutes without silane. Process 2 is similar to process 1, excepting that in addition to the pre-exposure of the reactor to silane, the annealing is performed under a silane flux varying between 0.5 and 5 sccm. In this study, both the duration and the temperature of the annealing have been varied, as can be seen in Figure 1 of supplementary information (SI).

The surface chemistry of the samples before and after annealing is systematically studied by X-ray photoelectron spectroscopy (XPS). First, a characteristic $Si_{2p}$ peak appears for processes 1 and 2 whatever the temperature and duration of the annealing (see figure 1 in SI) demonstrating a Si-surface enrichment. The optimum conditions for surface Si enrichment (highest $Si_{2p}$ peak intensity) are obtained in process 2 with 5 sccm of $SiH_4$ and at 1550°C for a duration between 5 and 15 minutes. Moreover, it has to be noted that the ratio of the $Al_{2p}$ peaks after and before annealing is always below 1 suggesting that Si atoms have substituted Al ones to form an AlSiN layer.

Several of these samples have been studied by cross-section TEM to determine the nature of the Si-rich surface layer. One sample has been specifically designed for TEM study and is studied extensively in the following: a MOVPE-AlN on sapphire layer has been annealed for 5 minutes at 1550°C with process 2 (only 0.5 sccm of $SiH_4$) and then transferred to a MBE reactor. A 280 nanometer-thick AlN layer has then been overgrown to protect the Si-rich layer during TEM sample preparation. TEM specimens are prepared using a conventional technique involving mechanical thinning followed by ion milling using Ar at 0.5–5 keV. TITAN THEMIS microscopes operated at 200kV and fitted either with probe or with objective corrector are used enabling spatial resolution below 0.1 nanometer. The probe corrected one is also fitted with a high sensitivity energy dispersive X-Ray (EDX) spectroscopy system. The AlN on sapphire samples are very insulating and therefore subject to charging effects during TEM observations. To overcome this problem, short exposure images have been acquired. The presented high resolution high angle annular dark field (HAADF) scanning TEM images are in fact the sum of 10 single images.

Figure 1 shows cross-section HAADF images of the Si-rich layer along 2 perpendicular AlN zone axes with their respective Fourier transform (FT). A characteristic contrast (observed for all annealed AlN layers as seen in SI figure2) is observed along the <10-10>$_{AlN}$ zone axis whereas no clear difference of contrast with AlN exists along the <11-20>$_{AlN}$ zone axis. The thickness of the layer with the characteristic contrast varies from 4 to 6 nanometers along the sample. In fact, the characteristic contrast begins at different depths from the surface (see red arrows in figure 1). The top part of this layer (nearly 2 nanometer) appears to have a lower crystallinity and a darker contrast. The FT of the <10-10> image reveals a triple periodicity in the films plane along the <11-20> direction and a double periodicity along the <0001> direction. On the other hand, there is no additional periodicity along the <10-10> direction as seen in the <11-20> FT image. This in-plane triple periodicity is also observed in plan-view selected area electron diffraction (see figure 2 in SI).

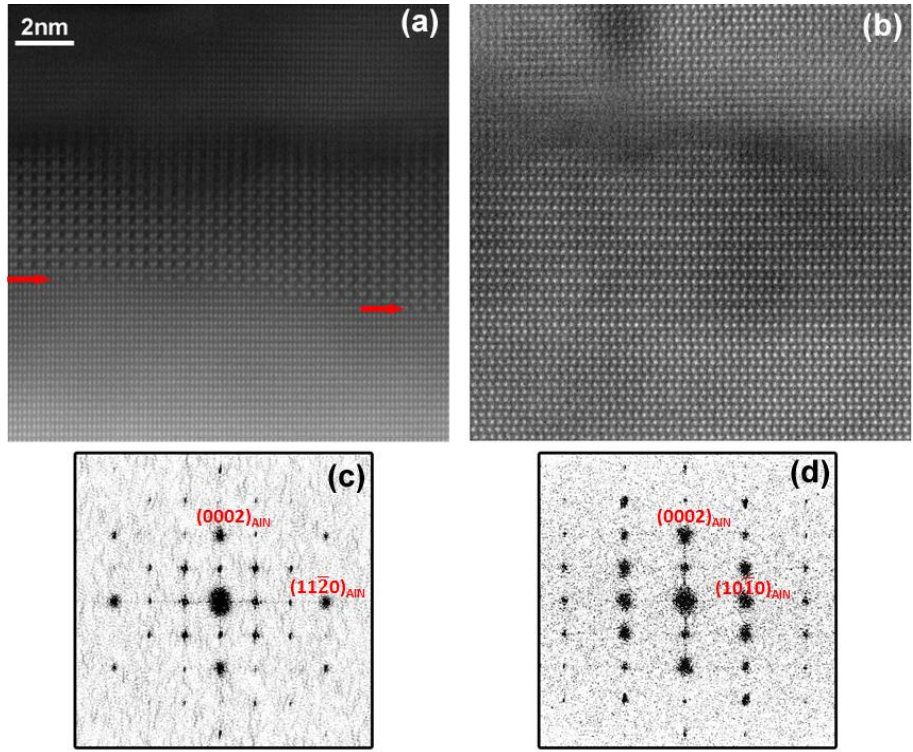

Figure 1: HAADF images of the Si-rich layer and their relative Fourier transforms: (a) and (c): <10-10>$_{AlN}$ zone axis; (b) and (d): <11-20>$_{AlN}$ zone axis. The red arrows point out the first plane showing the characteristic contrast of AlSiN.

The Si-rich nature of the layer is confirmed by EDX maps as seen in figure 2(a): the Al-content decreases as the Si one increases while the N concentration is nearly constant. Figure 2(b) is a quantitative profile across this layer showing the relative Al and Si concentrations. Coming from the substrate side, there is a gradual increase of the Si content up to about 50%. The characteristic contrast begins as soon as the Si-concentration reaches 50%. In the top part of the layer the Si-concentration increases above 50%. This region corresponds to the one with the lower crystallinity. A significant concentration of oxygen is also detected in this region (not shown) whereas no oxygen is detected in the bottom part. The sample surface has been exposed to air during the transfer from the CVD to the MBE reactors. It can be assumed that the O-enrichment is due to this exposure. The following investigation of the SiAlN layer structure will be focused on the bottom part of the layer (4nm) which has the better crystallinity and do not contain oxygen. From the data of figure 2(d), the mean Al and Si contents correspond to 47+/-4% and 53+/-4% of the occupied cation sites, respectively. The error bars correspond to the standard deviation from the mean values of the concentrations in the bottom 4nm of the Si-rich layer.

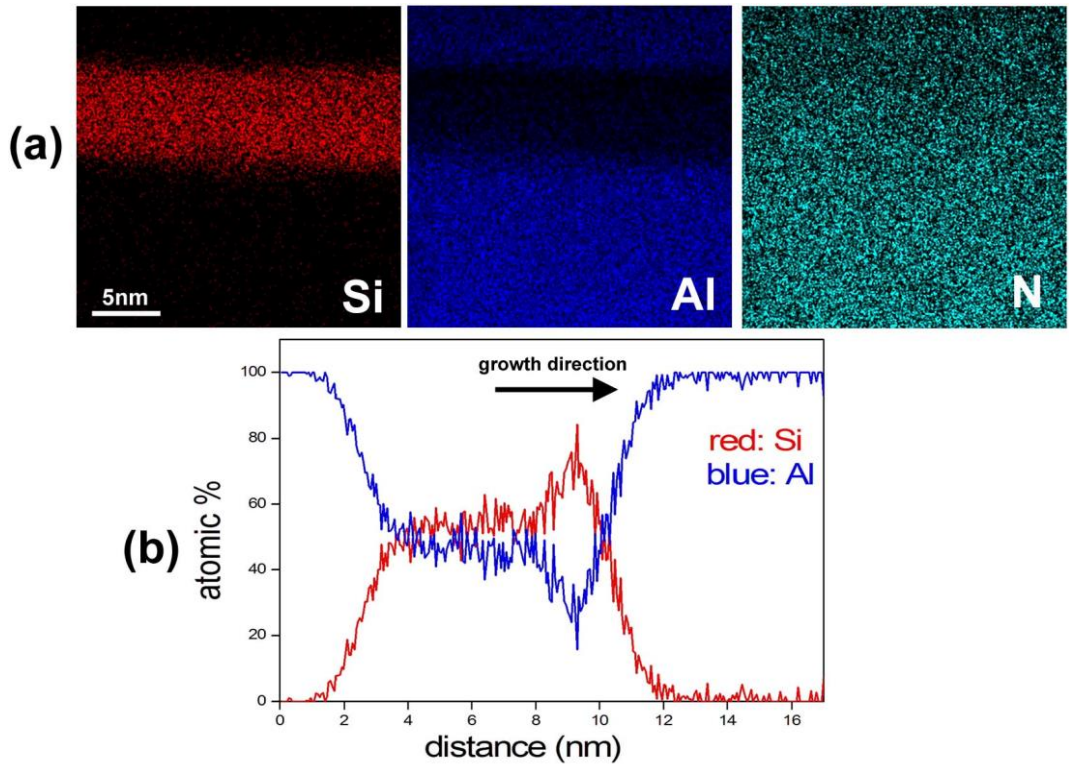

Figure 2: (a) Si, Al and N EDX map of a Si-rich layer; (b) Profile of the relative Al and Si concentration across a Si-rich layer

Figure 3(a) is a high resolution TEM image along the <10-10>$_{AlN}$ zone axis. The top part of the layer has been removed during the TEM sample preparation process. Figure 3 (b) and (c) are strain maps of figure 3(a) obtained using the geometrical phase analysis (GPA)[15] method for {11-20} ($\varepsilon_{xx}$) and (0002) ($\varepsilon_{zz}$) planes respectively. Figure 3 (d) shows the intensities in figure 3 (c) summed parallel to the interface. There is no strain for <11-20> planes showing that the in-plane lattice parameter of the new structure is equal to the AlN one. Hence, the new structure is fully coherent with its AlN substrate. On the other hand, the out-of plane lattice parameter is 3% smaller than that of AlN. The new structure being in coherent epitaxial relationship with the AlN template, the measured parameters may result from epitaxial strain.

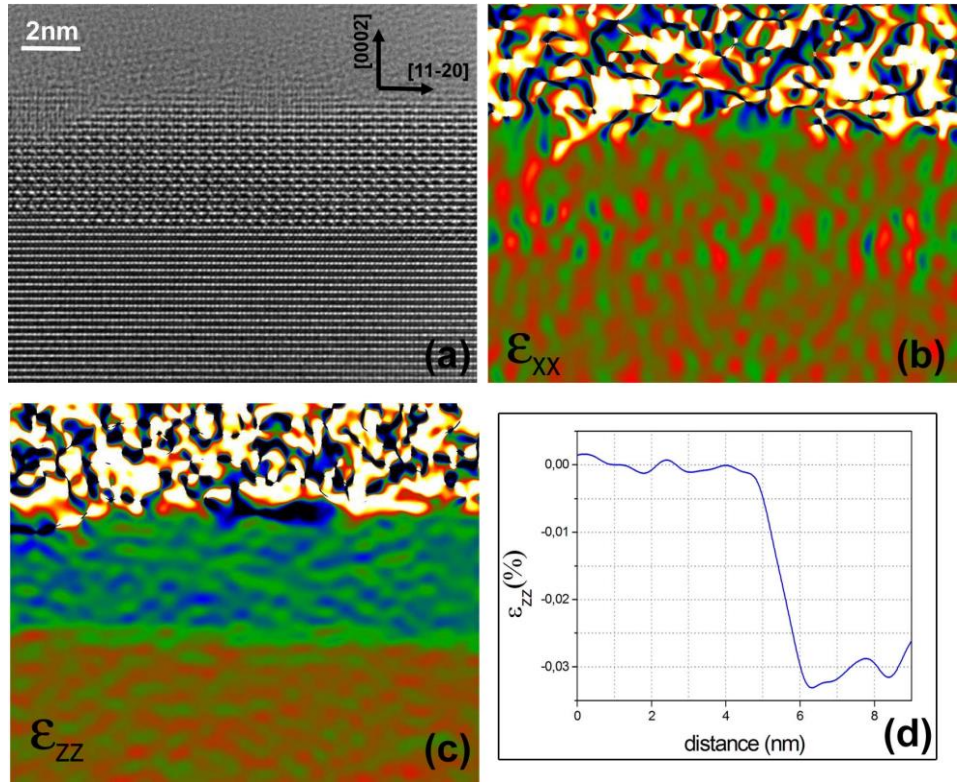

Figure 3:

  (a) High resolution TEM image along the <10-10>$_{AlN}$ zone axis.
  (b) In-plane GPA strain map extracted from (a)
  (c) Out-of-plane GPA strain map extracted from (a)

Figure 4(a) shows a characteristic high magnification Wiener-filtered HAADF image of the Si-rich layer along the <10-10> zone axis. The structure may be described as the stacking of 2 different planes along the vertical direction corresponding to the observed double periodicity. Plane 1 shows a homogeneous contrast of the atomic columns with intensities similar to the ones observed in the AlN template. However, a triple in-plane periodicity is clearly observed in plane 2. One of the three positions corresponding to the atomic columns of the AlN wurtzite structures shows almost no intensity and the intensities of the other two columns are similar to those of AlN and therefore to those in plane 1. The HAADF images have been acquired using a half-beam convergence of 20 milliradians, a camera length of 110 millimeters and a detector half-angle width extending from 65 to 200 milliradians. With such acquisition conditions, HAADF images show Z-contrast with atomic columns intensities proportional to $Z^{\alpha}$, Z being the average atomic number along the column and with α being between 1.6 and 2. Since Al and Si atoms have close atomic numbers (13 and 14 respectively), it is difficult to distinguish Al-rich or Si-rich columns.

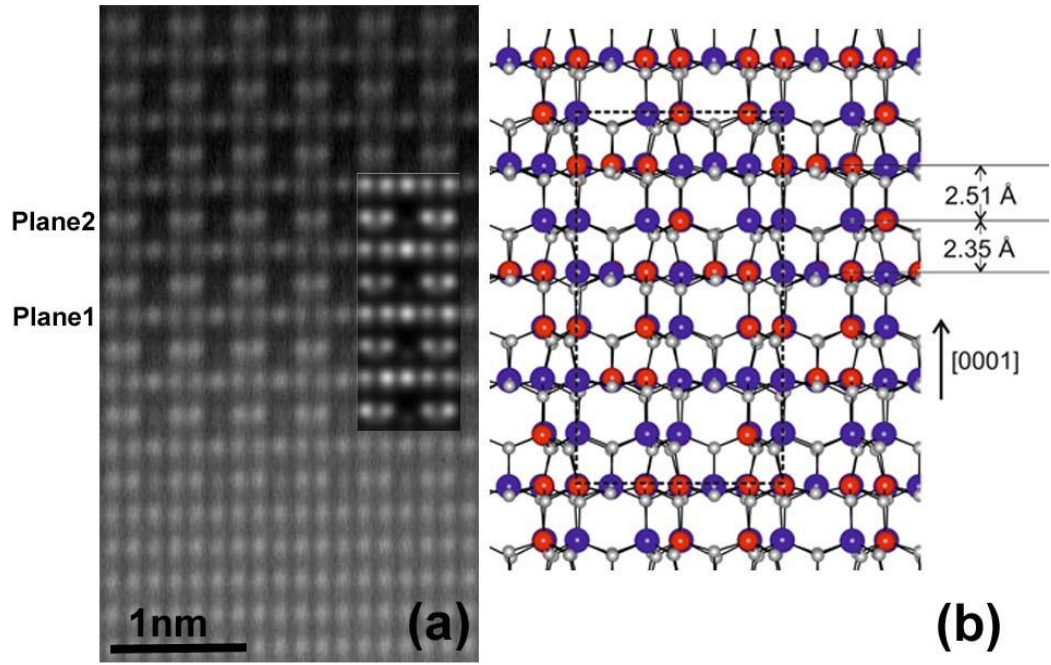

Figure 4:

(a) Wiener-filtered HAADF image of the bottom part of the AlSiN layer along the <10-10>$_{AlN}$ zone axis. The insert shows a simulated image using the model presented in (b).
(b) Schematic representation of the relaxed $Si_5Al_5N_{12}$ structure in the <1-100> projection. The dashed rectangle indicates the unit cell. Large red and blue balls denote Si and Al atoms, respective small gray balls the N atoms. The average interplane distances of the successive layers are also indicated.

From the above experimental results, a structural model can be proposed for the AlSiN structure. First, the compact hexagonal N-anion sublattice is assumed to be unchanged. Al and Si atoms are introduced in tetrahedral sites in the N-sublattice as in the wurtzite and $Si_3N_4$[16] structures. In plane 1, all atomic columns exhibit similar contrast in HAADF images and EDX indicates nearly the same Al and Si concentrations: the same number of Al and Si atoms is then distributed in the cation sites. Plane 2 presents a triple periodicity similar to the $GaSiN_3$ monolayer observed by Markurt et al.[6] allowing to expect a comparable structure. The empty cation position corresponds to the atomic column position showing no intensity in the <10-10> HAADF image. In the case of $GaSiN_3$, the two other cation positions are occupied by either a Si or an Al atom. As already be mentioned, it is impossible from our data to distinguish between Si- or Al- rich columns and/or ordering of these atoms along the <10-10> columns. Therefore, in the following discussion a random distribution of Si and Al atoms is assumed in the occupied sites of the cation sublattice.

Although the exact stoichiometry of the AlSiN structure cannot be obtained directly from the aforementioned experiments, under the assumptions that (i) the anion sublattice is fully occupied and (ii) the AlSiN structure is semiconducting, a good estimation can be made. If we assume that the actual stoichiometry is $Al_xSi_yN_6$ then in order to achieve a semiconducting

AlSiN layer the electron counting rule (ECR) should be applied, i.e. all bonding and anion dangling bond states should be occupied. The total number of valence electrons per structural unit in the $Al_xSi_yN$ layer is $n_e=3x+4y+30$, where the last term is the total number of valence electrons contributed by the six N atoms. Moreover, the total number of electrons required to occupy all Si-N and Al-N bonding states as well as all N dangling bond states is 8×6=48 (6 N-sites and 6 cation-sites in the unit cell). Therefore, obeying ECR leads to $3x+4y=18$. If we assume equal number of Si and Al atoms then $x=y\approx2.57$ and in this case ≈14.3% of the cation sites should be vacant. This is somewhat smaller than the 16.6% cation vacancies required if the dark columns in plane 2 consist of vacancies only. The later could be achieved only if y=3 and x=2, i.e. if the Si and Al contents in the $Al_xSi_yN_6$ structure are 50% and 33% of the cation sites, respectively which is not what is experimentally-observed. A few Al and/or Si atoms should therefore be present in the sites corresponding to the dark columns. $Al_{5+1/3}Si_5N_{12}$ and $Al_5Si_{5+1/4}N_{12}$ correspond to the extremes of Al rich and Si rich stoichiometries, respectively, which allows to globally verify ECR. In the first case, the Al and Si contents are 51.6% and 48.4%, respectively, and 1/6 of the sites in dark columns are occupied by Al atoms. In the second case the Al and Si content are 48.8% and 51.2%, respectively, and 1/7 of the sites in dark columns contain Si. It should be noted that both of these sets of contents are compatible with the contents determined by EDX. We can therefore propose a stoichiometry $Al_{5+\alpha}Si_{5+\delta}N_{12}$ with α being between 0 and 1/3 and δ between 0 and 1/4.

Density functional theory (DFT) calculations within the local density approximation (LDA) for the exchange and correlation and the projector augmented-wave (PAW) method have been employed in order to calculate the relaxed atomic geometry of the experimentally suggested structure of the AlSiN layers. The AlSiN layers have been modeled using supercells consisting of 8 N-cation ML along <0001> with a $2\sqrt{3} \times 2\sqrt{3}$ periodicity in the basal plane. The plane-wave energy cutoff was 450 eV and an equivalent of a 6×6×1 Monkhorst-Pack k-point mesh for the unit cell was used to sample the Brillouin zone (BZ). All the atoms in the supercells have been allowed to relax until the forces are smaller than 1 meV/Å. Furthermore, the AlSiN layers have been assumed to be biaxially strained to AlN, i.e. the in-plane lattice constant is fixed to the lattice constant of AlN and the supercells were allowed to relax the strain along the <0001> direction. A schematic representation of the relaxed structure is shown in Figure 4(b). The cation sublattice is a stacking of two alternating layers along the c-direction. Both layers contain equal number of Si and Al atoms randomly distributed. However, 1/3 of the cations sites in plane 2 are vacant. The vacancies are distributed in an ordered $\sqrt{3} \times \sqrt{3}$ pattern and the structures stoichiometry is $Si_5Al_5N_{12}$. The out of plane lattice constant is ≈ 2% contracted with respect to the corresponding lattice constant of AlN. This lattice contraction is in good agreement with experimental observations and the difference can be attributed to larger concentration of the cation vacancies in the calculated structure with respect to structures that obey ECR.

The relaxed structure has been used as an input for HAADF image calculation with the JEMS software[17] using the experimental parameters described above. The insert in Figure 4 (a) shows a simulated image for a specimen thickness of 5 nm. The good fit between the simulated and the experimental images allows to validate the proposed structural model.

Nevertheless, a few questions on the proposed structural model and stoichiometry have to be further addressed. Firstly, the vacant cation sites of the proposed structural model presents a triple periodicity in the film plane and a double periodicity perpendicular as compared to the wurtzite one of the AlN templates. Therefore, translational domains, such those formed when planes 1 and 2 are shifted by 1/2 $c_{AlN}$ in adjacent domains, cannot be excluded in the microstructure of the AlSiN layer. A shift of the vacant cation sites in adjacent domains may also be present. Secondly, we have assumed a random distribution of Al and Si on the cation sites. This assumption cannot be validated by HAADF experiments due to the very similar Al and Si atomic masses. However, disordered or ordered distributions of Al and Si atoms and/or small changes in the exact stoichiometry may have drastic consequences on the physical properties of AlSiN and should therefore be investigated. Nevertheless, these issues are beyond the scope of the present paper.

In conclusion, we have shown that the high temperature annealing of AlN epitaxial layers under a Si atmosphere allows to synthesize a new Nitride compound. Based on HRSTEM, EDX measurements and DFT calculations a structural model and a stoichiometry ($Al_{5+\alpha}Si_{5+\delta}N_{12}$) are proposed. Although, the electronic structure is not known at this stage, the parent materials, AlN and $Si_3N_4$, are both wide gap semiconductors. Therefore, we may speculate that $Al_{5+\alpha}Si_{5+\delta}N_{12}$ is also a wide gap semiconductor. A detailed investigation of the properties of this new material is nevertheless appealing both from a fundamental point of view as well as to assess potential applications.

# Supplementary Information

**X Ray Photoelectron spectroscopy**

The surface chemistry of AlN layers has been investigated by ex-situ X Ray Photoelectron spectroscopy as a function of the annealing conditions. Figure 1 gives the ratio of $Si_{2p}$ ($Al_{2p}$) intensities after and before annealing. The ratio is greater than one for Si (Si-enrichment of the surface) and lower than 1 for Al (Al-impoverishment of the surface)

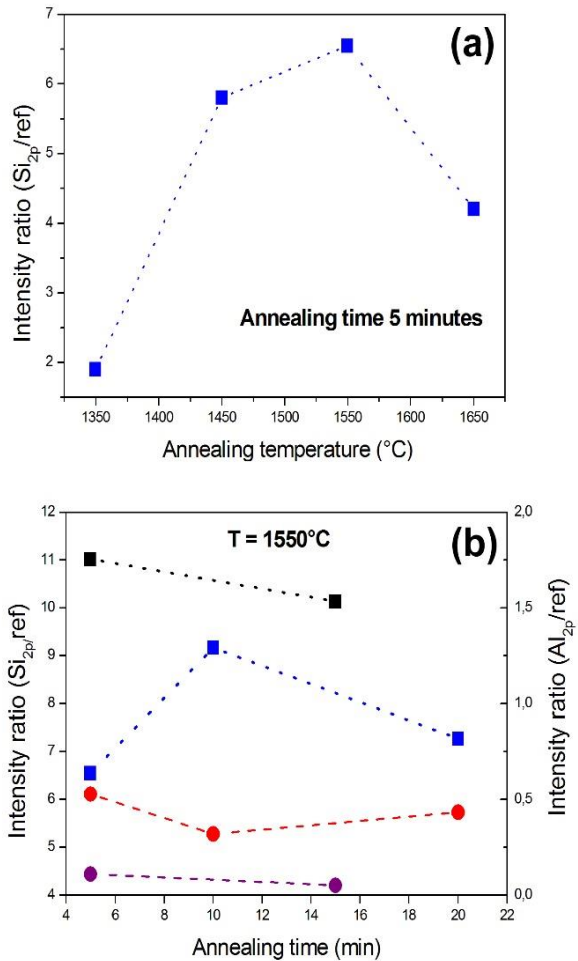

Figure 1 SI : ratio between the $Si_{2p}$ ($Al_{2p}$) peaks after and before annealing as a function of the temperature for a 5 minutes annealing (a) and as a function of the time for an annealing at 1550°C (b). In (b), the square symbols correspond to $Si_{2p}$ ratio (left scale) for process 1 (blue) and 2 (black) whereas round symbols correspond to $Al_{2p}$ ratio for process 1 (purple) and 2 (red).

**Surface AlSiN layer**

The paper is focused on the study of AlSiN layer embedded in AlN. AlSiN layers are also clearly detected by TEM even without an AlN overgrowth. Figure 2 SI shows the presence of such a AlSiN layer for an annealing of a MBE-grown AlN layer for 5 minutes at 1550°C with process 1. This sample has also been studied by plan-view TEM. The plan-view selected area electron diffraction pattern shows the triple periodicity of AlSiN along the <11-20> directions.

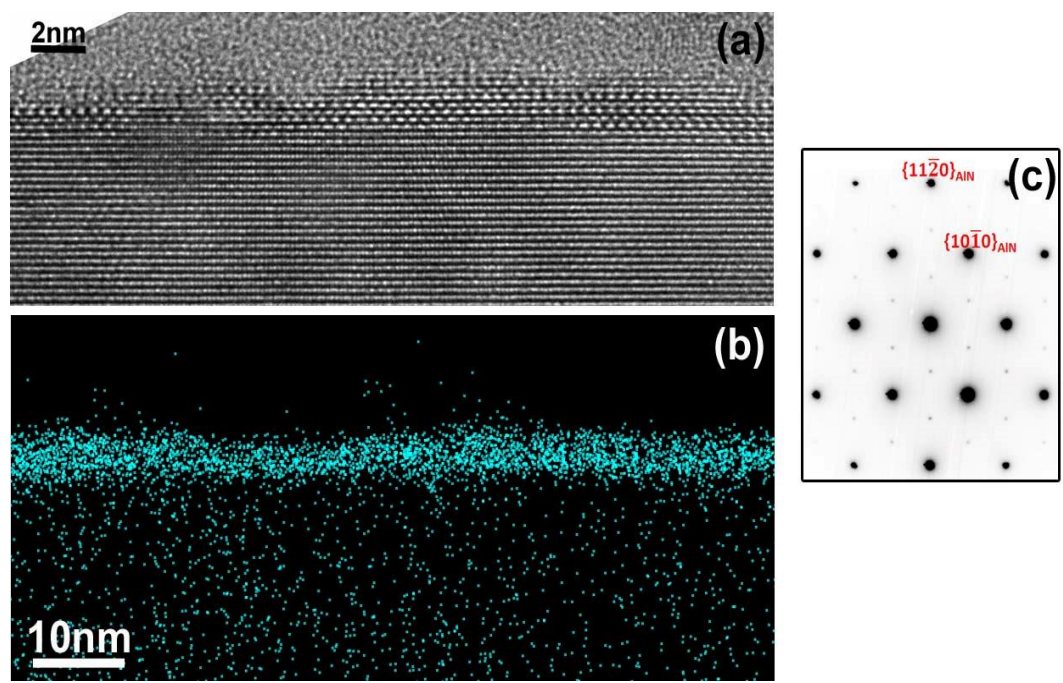

Figure 2 SI :

(a) High resolution TEM image of a MBE-AlN layer annealed at 1550°C with process1
(b) Si EDX map
(c) Plan-view selected area electron diffraction pattern. The in-plane triple periodicity is clearly observed.

**Acknowledgement**

The part of this work done on the NanoCharacterisation PlatForm (PFNC) was supported by the "Recherches Technologiques de Base" Program of the French Ministry of Research. PV want to thanks CP2M Marseille for the access to the objective-corrected TITAN microscope and to T Neisus (University P. Cezanne, Marseille) for his expert help in the use of this microscope.